# The perceptions of social and information privacy risks of Inflammatory Bowel Disease patients using social media platforms for health-related support

Kate O'Leary, Elvira Perez Vallejos, Neil Coulson, Derek McAuley

**Abstract**

With hundreds of thousands of individuals using social media to discuss health concerns, sensitive personal data is self-disclosed on these platforms every day. Previous research indicates an understanding of social privacy concerns by patients with chronic illnesses, but there is a lack of understanding in the perception of information privacy concerns. Qualitative interviews were conducted with 38 patients with inflammatory bowel disease (IBD) using social media platforms to engage with online communities. Using thematic analysis, we identified that patients typically demonstrate the privacy and risk dual calculus for perceived social privacy concerns. Patients demonstrate mixed knowledge of what data is collected and how it is used by social media platforms and often described their platform use as a trade-off between the unknown information privacy risks and the therapeutic affordances of engaging with the online community (the privacy calculus). Our findings illustrate the different levels of understanding between social and information privacy and the impacts on how individuals take agency over their personal data. We conclude with the suggestion for future research to further understand the relationship between knowledge, information privacy concerns and mitigating actions in the online health community context.

**Acknowledgements**

This research was supported by the University of Nottingham (RCUK Grant No. EP/L015463/1).

**Introduction**

Nearly half the world is connected through social media (We Are Social, 2018) and everyday billions of data points are collected, stored and processed by these platforms. There is growing concern over the obfuscated practices of online companies, specifically social media platforms that profit from explicit and implicit user-generated content (Crawford & Schultz, 2014). Despite concerns raised by researchers and news media outlets, people continue to use these platforms. The reasons for continuance intention are manifold including, but not limited to, people having a sense of belonging (Lampe et al., 2010), platform functionality (Lankton & McKnight, 2011), entertainment (Lampe et al, 2010) and addiction (Joo et al., 2016). For patients with long term health conditions, social media platforms serve as a meeting place for socio-emotional support and information (Coulson, 2015).

By law, health data is considered as one of the most sensitive types of information (GDPR, n.d); in online health communities (OHC), patients may disclose experiences with symptoms, medications, mental health, and surgical interventions as a result of dealing with an illness (Coulson, 2013; Roccetti et al, 2015; Britt, 2017 Matini and Ogden, 2015). Self-disclosed health information on social media platforms is less regulated than in the healthcare context; meaning that the information can be more freely processed and shared (Patterson, 2013). For reference, Facebook's data policy indicates that information may be shared with third-parties to advance innovation in areas such as health (Facebook, 2018).

Privacy and its definitions are often positioned around the control of information flows through different contexts. Individuals assess the risk of sharing information based on the nature of the information, who it is shared with (and their role in that context), as well as how the information is transmitted (Nissembaum, 2004). This paper explores the privacy concerns held by individuals with a lifelong chronic illness Inflammatory Bowel Disease (IBD). We outline two different types of privacy: social privacy and information privacy, explaining their variance in the context of online health communications. We share our findings from an interview study through the dual lens of social and information privacy, observing how they are perceived and acted upon in different manners.

**Background**

Inflammatory Bowel Disease affects over 300,000 people in the UK (Crohn's and Colitis UK, n.d.), 1.5 million people in the United States (Crohn's and Colitis Foundation of America, n.d.) and thousands more worldwide. Crohn's Disease and Ulcerative Colitis are the two main constituents of IBD (Hanauer, 2006). The hallmark of this chronic auto-immune disease is gastrointestinal inflammation causing symptoms including diarrhoea, abdominal pain, ulcers, nausea, intestinal bleeding and fatigue (Hanauer, 2006; Crohn's Colitis UK, nd; Head & Jurenka, 2003). Patients with IBD undergo a lifetime of care, experimenting with different medical interventions from anti-inflammatories to immunosuppressants, steroids and biologics, and in many cases, invasive surgery (Hanauer, 2006; Crohn's and Colitis, 2017d). The stigmatised non-visible nature of the illness makes empathy from people who do not live with IBD challenging (Hall et al., 2005; Wright & Bell, 2003; Seres, 2014); often causing patients to feel isolated from their day-to-day activities and relationships (Lönnfors et al., 2014; Jedel et al., 2016). Patients' work and school lives can often be interrupted as a result of their symptoms, hospital visits and surgery (Lönnfors et al., 2014; Liu, Inkpen & Pratt, 2015; Carter, Qualter & Dix, 2015).

Patients have been accessing peer-to-peer support for Crohn's Disease and Ulcerative Colitis on the internet since the 1990s (Crohns Forum, n.d.). From bulletin boards and web-based forums, patients have adapted their social media use to encompass their health-related support needs. Patients seek and share their concerns and experiences about their illness, self-disclosing information about their symptoms, medications and day-to-day life (Coleman, Shah, and Jain 2015; Turner, 2017; Lewinski et al., 2017; Coulson, 2013; Coulson, 2015a). Accessing social support can improve an individual's quality of life (Moskovitz et al., 1999) with IBD, with Coulson (2015) identifying that accessing support online is not only more convenient than face to face, but has been reported to improve patients' wellbeing (Coulson, 2013; Coulson, 2015; Summers, 2018).

As individuals share tremendous detail about their health on the internet, that in a medical context the handling of such information is strictly regulated. How individuals perceive privacy in a non-medical environment offers an interesting space for research. Although everyone has their own personal preferences towards privacy (O'Hara, 2016), it has been seen that people behave in ways that contradict their privacy concerns. This is described as the 'privacy paradox' (Lutz and Strathoff, 2011; Taddicken, 2014). Often the study of privacy through computer-mediated communications has centred around the willingness to self-disclose. However, when considering privacy through the lens of contextual integrity (Nissembaum, 2004) (which contends that privacy boundaries are formed by societal norms, the law, and personal preferences) disclosure serves as one facet of privacy management. Instead, a privacy violation occurs when the context is compromised in some way. This is otherwise known as a 'context collapse' (boyd, 2014).

Media scholars contend that the majority of individuals in online communities participate peripherally or 'lurk', suggesting that most do not actively self-disclose (Jenkins, 2013). Non-disclosure in health communities however, is not entirely indicative of a high privacy concern. For instance, patients who are interested in learning information about their illness are less likely to self-disclose, while those seeking emotional support are more likely to post online (Nonnecke et al., 2006; Welbourne et al., 2012; Choi et al., 2017). Despite previous, albeit limited, research showing that individuals with a health condition have increased feelings of self-stigma (Liao, 2019), and anticipate social privacy risks (Patterson, 2013), self-disclosure is higher amongst those with poorer health (Liao, 2019; Frost et al, 2014; Zhang, 2018).

When addressing notions of context collapse within online health community research, patients have demonstrated an awareness that the information they post online may be read by other unintended audiences (Patterson, 2013; Newman et al., 2011; Brady et al., 2016' O'Kane et al., 2016). Concerns included: threats to employability; discrimination; fear of stigma and being judged; impact on personal relationships; and fear of being hurt by others (Naslund & Aschbrenner, 2019; Moore et al,

2016; Patterson, 2013; Zhang et al, 2018; Newman et al., 2011; Brady et al., 2016). When individuals recognise what their concerns are, it enables them to weigh the benefits against the risks. This is known as the privacy calculus (Zhang et al., 2017). However, previous studies have identified how patients who engage with health communities on the internet, have developed behaviours to mitigate risks of context collapse. Patients may choose to censor the information they share so that when a context collapse occurs, such as during a job interview, they feel they have the confidence to explain themselves (Brady et al., 2016). Patients will also choose what information to share with different networks; such as sharing personal experiences with health communities but not with family and friends (Newman et al., 2011; Brady et al., 2016).Meanwhile, some feel very uncomfortable with sharing information with strangers (O'Kane et al., 2016). The comfort in sharing experience may change over time as contexts change. One participant from Brady's study felt 'too vulnerable' to share their experiences of fertility treatments until after she became pregnant, "indicating that her desire for privacy was shaped by the need to control the context in which the information was shared rather than the information itself," (Brady et al., 2016, p5). The net privacy risk after individuals take mitigating actions is lower than those who do not (Zhang et al., 2017). The privacy calculus and the risk calculus are applied together by patients who choose to disclose health information on the internet (Zhang et al., 2017).

Despite individuals in OHCs reporting to have capabilities in perceiving and mitigating privacy concerns pertaining to disclosure of information to unintended audiences, privacy researchers, journalists and activists describe a different set of concerns about how online platforms collect and process personal information. Personal and behavioural data is collected on mass (Big Data) and processed by algorithms to profile individuals. Unanticipated by users, personal data (and algorithms) can be used for targeted advertising (Speicher et al., 2018), content filtering (Barker, 2017), emotional grooming (Kramer, Guillory & Hancock, 2014), political manipulation (Dutton et al., 2017), and predicting depressed users (Fowler, 2012; Eichstaedt et al., 2018). The unintended impacts of algorithmic decision making about individuals may include 'personalised' health insurance premiums (Crawford and Schultz, 2014; Scism, 2019). With the rapid developments in predictive analytics and their integration with real world outcomes, these unintended outcomes are constantly changing. It is also worth noting that real-world contexts are also constantly changing; for instance, healthcare in the United States changes between administrations (Rice et al., 2018) and the increasing privatisation of public services in the UK (Guy, in press). To date, the evidence of the long-term impacts of personal data processing is anecdotal. The asymmetry of knowledge as a result of secret 'black box' algorithms means that neither researchers nor users can predict the impacts, when they will occur, who they will occur to, and to what degree. These impacts are described as unknown unknowns (Wilton 2017; Rumsfeld, 2002). In OHC research, there are few studies that explore similar privacy concerns shared by privacy advocates. In a study on health apps and sharing information to social media platforms, participants described concerns with health insurance eligibility and targeted advertising (Patterson, 2013). The author of this paper (Patterson, 2013) echoes how these privacy risks are more difficult for individuals to identity and mitigate. This is said to be as a result of a low understanding of information flows, which are unclear in terms and conditions (Rao et al., 2016; Debatin & Lovejoy, 2009; Lutz & Strathoff, 2011).

If personal data is shared between other individuals, and also the platform provider, it is important to distinguish these two types of information flows and the privacy implications therein. Lutz and Strathoff (2011) differentiate privacy into two types. Firstly, social privacy concerns "describe the fear of intrusion caused by other people" (Lutz & Strathoff, 2011, p85). Meanwhile, concerns over health insurance eligibility (Patterson, 2013; Crawford & Schultz, 2014) and unwanted targeted advertising (Patterson, 2013) evoke information privacy concerns (Lutz & Strathoff, 2011). These privacy concerns delineate people's discomfort with institution that use their data for unwanted purposes.

Despite a comprehensive quantitative study (Zhang et al., 2017) indicating how mitigating actions negatively influence net privacy risks, there are few qualitative studies that explore contextual integrity and privacy concerns among people sharing health-related information. Current studies demonstrate how patients negotiate social privacy concerns through risk mitigating behaviours (Patterson, 2013). There is little in the way of how, and indeed if, patients perceive information privacy concerns when sharing information about their health on contemporary social media platforms. This study explores the risks that are identified by patients with Inflammatory Bowel Disease through the perspectives of both social and information privacy.

**Methodology**

This study received ethical approval (CS-2017-R5) in January 2018 to recruit participants by the Computer Science Ethics Committee at the University of Nottingham. Participants received an information sheet and a consent form prior to their involvement in the study and were given opportunities to ask questions as well as withdraw.

Semi-structured interviews were conducted through video conferencing (n=15), over the telephone (n=22) and face to face (n=1). Participants were patients who have IBD and use social media platforms to engage with an online health community relating to the illness. Participants were asked a series of questions regarding their preferences over their self-disclosed information. Questions were framed around the notions of the privacy calculus and context. Participants were invited to share their experiences of engaging with online health communities as well as the perceived risks of sharing health-related information on the internet. They were asked about their views if the information was available to other audiences from both a social (family, friends, potential employers) and an information perspective (commercial entities and other third-parties). Participants additionally shared their understanding of what data is collected by social media platforms as well as how that information is used. Semi-structured interviews offer the flexibility for the researcher to ask follow-up questions to clarify points made by participants as well as provide rich descriptions of the participant experiences.

Posts were created and shared across the social media platforms Facebook, Twitter and Instagram to recruit participants with IBD. To better target potential participants, relevant hashtags were applied on Twitter and Instagram, while permission was granted by two Facebook Groups to advertise within the closed environment. Participants were offered a £10 voucher to compensate for their time and support. Patients with IBD are distributed all over the UK and the world and so study participants were invited to interview over the telephone or through the internet on Skype. This method enables participants to find a space comfortable for them to self-disclose, such as in their own homes (Hanna, 2012; Seitz, 2015). The interviews were audio recorded and transcribed by the first author (n=33) and a third-party transcription service (n=5).

Patterns within the data-set were identified through inductive analysis, which is "a process of coding the data without trying to fit it into a pre-existing coding frame, or the researcher's analytic preconceptions. In this sense, this form of thematic analysis is data-driven," (Braun & Clarke, p12, 2006). Codes pertaining to knowledge, attitudes and actions were grouped into broader themes of social privacy concerns, information privacy concerns, social privacy mitigations, trade-offs and information privacy mitigation.

A set of four transcripts were sent to the second and third authors to evaluate the inter-coder reliability in order to assess whether similar themes were identified. This process was conducted early in the thematic analysis procedure to ensure consistency of the remaining data analysis by the primary coder. Given the exploratory nature of the interviews, similar themes were identified for full analysis. All transcriptions were coded through NVivo; a research software package that facilitates qualitative data analysis.

**Findings**

In total, 38 participants (aged 16-70) with Crohn's (n=25), Ulcerative Colitis (n=12) and indeterminate colitis (n=1) were interviewed for this study. The split between male and female participants was near even with 18 and 20 participants respectively. Participants were residents in the UK (n=32), USA (n=5) and South Africa (n=1) and described their experiences in English speaking OHCs. Participants demonstrated a clear understanding of social privacy risks, however, there was a lot more uncertainty around information privacy concerns. The following findings explore: the identified social privacy concerns; mitigating actions; knowledge of how personal data is processed by social media platforms; and information privacy risks.

*Social Privacy Concerns*

Synonymous with Patterson (2013) and Brady et al., (2016), a fifth of interview participants with IBD expressed a concern for their career progression as a result of online surveillance. The online vetting practices demonstrated by employers was considered to be a *"disgusting"* (Participant 37) process that should not be used to determine whether someone is appropriate for a job. Other participants were less concerned; trusting that employers will not use this information to unfairly discriminate against their eligibility. It could be inferred that there are other antecedents to this privacy concern including previous experiences with employers as well as employment status. Individuals who described themselves as being well established in their careers, or even retired, perceived a much lower risk of career harms as a result of online vetting by employers.

> *"I retired early because I thought I might as well retire when I can rather than keep working with the Crohn's so it doesn't affect me work wise but for other people where I am told employers now do as part of a normal search a social media search on you I can see that could affect someone's employability." (Participant 5)*

Many participants, but not all, described discomfort with self-disclosing health-related information with family and friends. Some perceived this information to be too upsetting for family members while others are concerned that a lack of empathy from non-sufferers may give way to negative feedback.

> *"I don't want certain people to know about certain things so I am very hesitant sharing stuff on my public Facebook account just because I don't want any backlash from it, not that there would be or anything like that but it's just, it's just somewhere I don't really share things like that." (Participant 1)*

> *"I put on Facebook "yeah not feeling very well" or "just got home in time before my guts exploded" that kind of thing but Twitter is where I really put "I can't really go on today I need to just curl up" because my mum's not on there and she doesn't really understand; she just thinks I've got an upset stomach even now. And I think if I keep putting that on she goes 'ooh you share too much with everybody' even though she has a habit of oversharing. You know 'we don't always want to hear that you've got an upset stomach' so Twitter is my safe space." (Participant 11)*

When broadcasting to and communicating with strangers on the internet, there is no guarantee that everyone will be civil, despite the empathetic nature of support communities. Notions of 'trolling' and receiving hurtful comments from other internet users was identified as a risk for six of the interviewees, but something that should be expected and prepared for.

> *"I mean the only thing I can think of off the top of my head is maybe derogatory comments which is purely what the internet can be. Keyboard warriors if you will but no I think that the rewards far outweigh any risks."* (P20)

Similar to previous studies by Naslund & Aschbrenner (2019), Moore et al (2016), Patterson (2013), Newman et al (2011), and Brady et al (2016), the IBD patients represented in this study identified similar social privacy concerns. Participants shared varying levels of concerns around disclosure to real life audiences as well as receiving unkind comments from strangers on the internet.

*Social Privacy Mitigations*

While not all unintended audiences reading self-disclosed health information will amount to negative outcomes, participants reported techniques they employ to better control the context in which they discuss their health. Similar to studies on how young people preserve their privacy on social media (Marwick & boyd, 2014), participants employ techniques such as audience management and self-censorship to control what information is shared with different audiences. Participants managed their audiences through using multiple accounts on one platform, such as creating a separate account to disclose health information and participate in the IBD OHC, thus enabling them to filter their audience. Participant 17 described having two Twitter accounts as *"it's better for me to talk about my IBD through my own personal one rather than having it go into my work account"*. Participant 11 described using Twitter as a space to talk about their symptoms and experiences with treatments, saying that *"my Twitter is my safe space; also there are not a lot of work colleagues on there"*. Participant 26 said that it's *"weird"* they don't share on Facebook because *"obviously they're the people I actually know, but there's a lot of professional relationships on there with people that I actually work in the same office with"* and they don't want to have people in their real life questioning them about what they disclose about their health, which was intended for a different audience.

In spaces that individuals have identified as safe for their personal privacy preferences, self-disclosure increases. For instance, individuals who use Facebook Groups reported how posts are more *"graphic"* and *"personal"* than in other environments. Participant 2, who created their own Facebook Group and also uses Twitter and Instagram described that if: *"[a group has] a little more privacy to it people are more inclined to share a little more than they would normally. Or ask questions that you know on a different forum they might be criticised for or made fun of"*.

Participants demonstrated self-awareness of their privacy boundaries, identifying what information they were comfortable with sharing with different audiences. In order to still engage with the online health communities across Facebook, Twitter and Instagram, participants described nuanced tactics to preserve the contextual flows of their information; reducing social privacy risks.

*Information Privacy: Understanding of the Digital Economy*

Corresponding with previous research that indicates extremely low engagement with terms and conditions, only one participant in this study reported to have read them. Their reasoning for reading them is that their job in cybersecurity warrants a strong understanding of what is delineated in the policies. The majority (n=31) did not read the terms and conditions at all, while a small number (n=6) reported to have skim read them. Given these findings, knowledge of the empirical facts was expected to be low; therefore, the following findings on participant understanding of data collection and processing, should be reviewed as perceived knowledge.

The knowledge of the data that is collected by social media platforms varied; however most of the participants described a low level of understanding of what is collected, broadly stating that the platforms collect 'everything' without remarking on any specifics.

> *"I think everything but my first born! I don't know I mean I think they collect everything that you do."* (P10)

Synonymous with data collection, there is a lot of uncertainty expressed by participants over what happens with the information collected by social media platforms. Some participants "*haven't really thought of that*" (P1) until being asked in the interview. However, without clear knowledge of what data types are collected, many participants described a connection between their behaviours and targeted advertising that appears on their social media feeds. Their observations have led them to believe that these behaviours are collected in some way for advertising purposes.

> *"Erm they definitely know my like I guess my browsing habits and the things I click on within Instagram and I think even outside of Instagram and I'm personally not entirely sure exactly how it works I know there have been, I've seen ads for products that I've been looking at online or have recently purchased and it's like it's really obvious the connection they're making but yeah I mean I don't see them as doing anything particularly nefarious with that data. For every intensive purposes I know that they definitely follow me around."* (P16)

Three participants shared their observations between behaviours and adverts. However, they demonstrated an awareness that these behaviours are collected by cookies; small text files that are downloaded into the browser by first and third-parties. This understanding demonstrates a more sophisticated level of knowledge about data processing by online platforms. A small minority (n=2) of participants were more knowledgeable about data collection, describing the different data types that are gathered by social media platforms. Both participants described their knowledge from their involvement in the tech-industry, in advertising, and in cyber-security. Therefore, while some patients do have more knowledge, it appears to be career specific.

> *"Let's see we've got: location, age, we have pretty much any demographic, sex, gender, interests, accounts you follow, when they're online, they have your location, they may have you payment information, they have who you're connected to, I can go on for quite a long time" (P30)*

Beyond processing data to target adverts, some participants believe that their data is being directly shared or sold to third-parties; however, none of the participants could report who the third-parties were.

> *"I can yeah I can tell you that they sell information for profit and I believe that there are clauses in the Facebook license agreement, sorry that's not the correct expression, the use of terms, that says they can do whatever the hell they want and everyone signs up to it." (P34)*

Some participants were aware of their low understanding of what information is collected and how it is used by social media platforms. They said that following the interview they would read more into it so that they were more informed on the topic.

> *"What do they do? I really want to know now. […] I need to research this stuff actually, it's really annoying me." (P1)*

*Information Privacy Concerns*

Like social concerns, information privacy concerns varied according to personal preferences. For instance, while participants remarked that advertising is a normalised part of social media, they are perceived differently by the IBD community. Some participants described the usefulness of targeted advertising, improving their experience in a more personalised way. This personalisation to others is discomforting and *"intrusive"* (P38), making individuals feel as though they are being watched by companies, with participant 13 reporting that: *"It is a bit scary like 'big brother's watching you' a little bit yeah"*. As many participants had not previously thought about how social media platforms collect and use information about them, when advertising was discussed it appeared as though people *"don't really worry about the greater effect of it, in terms of collecting health information"* (P38).

A few of the participants remarked on commercial companies having access to information that patients have self-disclosed on social media platforms. They recognised that the information they share may be of value to some companies' research and development, however participants questioned a company's legitimate interest in serving patient needs: *"I get it because it's like they're looking up their end users, but at the same time these people are people with serious health conditions and they're trying to make money off of them at the end of the day"* (P19). While many participants were happy with supporting research, many shared the attitude that *"I don't think that anybody that hasn't been given permission by yourself should be sharing your information or any of your illness with people you're not aware of getting it"* (P31).

Although there are some concerns around how companies use their data, particularly when it is without consent, some participants felt as though they were not going to be personally harmed as a result of data collection and processing. Participant 32 said: *"I think if [big corporations] can make a sale then they will so I don't feel my information is safe and secure on the Facebook thing […] but I do feel a bit protected because I do feel like it's a barrier between them and me, the internet that is, it's a barrier, so I don't feel like I'm personally in any danger"*.

When asked about the potential risks of sharing health information on the internet, one participant from the UK described their concern with government surveillance and the risk that their online profile may negatively impact on their eligibility for state support.

> *"You need to be careful about saying certain things if you're saying it on your profile and not in a closed group but then other people could be on a closed group because I do know that PIP assessors have looked on people's profiles to see how active they are if they've done this, that or the other and I don't know if they're that sneaky that they would go onto closed groups or not I don't know, so I've become more mindful of that aspect but only that really. …"* (P24)

*Trade-offs*

When participants were asked to consider their perceived risks from sharing information about their health on the internet, many of them described a trade-off between receiving support from the community at the risk of their information being used by social media companies. Even when trust is breached by social media companies, such as with the Cambridge Analytica Facebook scandal (Granville, 2018) that hit headlines during the interview recruitment, participants described an intention to continue using the platform despite these risks. Participants demonstrated an implicit use of the privacy calculus; when the benefits were weighed up against the perceived privacy risks.

> *"It does feel a little uneasy but then I think there's loads of people that do it and if I don't take that risk I won't get the support from these groups. Swings and roundabouts really"* (P32)

The notion that personal data is the cost to access and engage in social media platforms (and therefore many online health communities) is synonymous with viewing data as an economic asset. Participants were cognisant that the services are 'free' and that social media platforms have to make money in some capacity, as Participant 5 explained: *"If I'm going to use them for supposedly free, the cost is my data. Which they will then go and make huge amounts of money"*.

Even though medical data is considered as sensitive information, participants struggled to identify the value of their self-disclosed information to social media platforms, suggesting that platforms "*must be pretty blummin' bored to collect my information; the ramblings of a 44 year old" (P11)*. There was a common perception that platforms would be able to make inferences based on their posting behaviours with great difficulty as *"it would take a lot of research to extract the information that would be useful to them, I think. From what I've put out there. I don't think the benefits to them of what they'd get out of it would be worth the cost of the research that they would have to do"* (P5). Previous research has demonstrated that individuals financially rate their personal information more highly when they have a market awareness of who else is interested (Spiekermann & Korunovska, 2017). When social media users are unaware of the full context within which they are sharing information, potentially with third-parties, they are more likely to devalue the data they have shared.

*Information Privacy Mitigations*

In addition to the social privacy settings available to users, the participants were asked whether they have consciously adapted their behaviours to mitigate information privacy risks. Unsurprisingly, the majority of participants claimed that they had not done anything beyond the privacy settings available to them on each platform. A significant finding however, is that a small number of participants (n=2) believed that the social privacy controls of Facebook Groups should mean that the platform should not use this information; to preserve the contextual integrity of the information and members.

> *"I trust that they [Facebook] won't [share information] because they're closed forums, but now you've made me think. I might go and do some research on that after this conversation." (P28)*

While there is a risk that social privacy settings are conflated with information privacy mitigations, some participants voluntarily explored their device settings during the interview, discovering settings they were hitherto not aware of. This may indicate that settings are difficult to find, as well as individuals not having enough knowledge to justify trying to mitigate against any information harms,

> *"Yeah I'm now going to do it, I'm going to look- ah I'm actually in account settings and oh location services is on, so I'm going to turn that off and now I've just done it. It sounds really stupid that even though I don't like it it's never really occurred to me to look into it." (P24)*

Two participants with computer science backgrounds described the additional use of third-party services to obfuscate behavioural information such as IP addresses and tracking cookies. Participant 38 described using *"Privacy Badger and others. […] What else? Well, I've got a VPN that I've got on my laptop and my phone,".* Meanwhile, a minority of participants described sharing minimal information with social media platforms, demonstrating self-censorship techniques, usually as a measure to mitigate financial fraud.

> *"There's information I won't put up there. I won't put up my mothers maiden name obviously, I wont put a true date of birth because what does every criminal want? Date of birth, maiden name and all that sort of stuff so I'm a little bit careful of what goes up." (P5)*

**Discussion**

Information privacy has been increasingly studied over the past few years (Kezer et al., 2016). Previous research has explored patients' social privacy concerns and behaviours online, finding that users typically self-censor posts and manage their audiences (Newman et al., 2011; Brady et al., 2016) to mitigate risks to their professional and personal lives (Patterson, 2013; Brady et al., 2016). In this study, participants described similar social concerns about the impact of different audiences viewing information about their health. While everyone has their own personal preferences over what information they're interested in restricting, participants described concerns with career progression; upsetting family and friends; annoying family and friends; receiving unkind comments; and personal safety. Even though some participants have accepted that they cannot control other people's actions, such as those who post cruel responses, the majority of participants reported risk mitigating behaviours in order to exert control over the context of their health information.

The identification of risks and their weighting against the perceived benefits of participation in online health communities is indicative of the privacy calculus. However, given the high level of mitigating actions, of self-censorship and audience management, to control the context of information, the risk, benefits and mitigations are more closely attributed to the 'risk calculus'. When participants demonstrated limiting the flow of information, such as with tightened Facebook privacy settings, they demonstrated a lower risk to threats such as employer vetting strategies. The understanding of the risks, coupled with the knowledge of how to mitigate them, provides individuals with a sense of self-efficacy to better control their social media information.

The majority of the participants (n=31) declared that they did not read the terms and conditions of the platforms they use at all, before signing up and disclosing information about their health. Without knowledge of the information privacy boundaries set by the platforms, participants demonstrated a low understanding of what information is collected and how it is used by social media platforms. Terms and conditions are known for being difficult to understand (Rao et al, 2016) therefore their ability to inform internet users of the respective platforms' operations is questionable. Most of our participants were unable to articulate what information social media platforms collect, however, there was a general understanding that the information is processed in some way to influence targeted advertising.

Longer term risks, such as the impact on health insurance (Patterson, 2013; Crawford & Schultz, 2014; Scism, 2019), were not identified as a concern amongst these participants. There are two explanations for this lack of foresight: firstly, the majority of the research participants in this study were based in the United Kingdom where healthcare is currently a freely available service. Secondly, potential consequences of health-disclosure on the Internet are difficult to predict and are therefore often over-looked as a 'true risk' that requires action (Hallam and Zanella, 2016). As participants were asked about what happens with their data on platforms and subsequently where that data goes in a wider online context, they were inquisitive and some explored additional settings on their devices during the interview.

The majority of participants had not described any mitigating risk behaviours to reduce data collection by social media companies. Participants described the trade-off between unknown long-term risks for short term benefits, such as support. This is indicative of the privacy calculus when the benefits outweigh privacy risks. In this case, the 'risk calculus' does not apply for the majority of the participants because they did not demonstrate mitigating actions.

With so many participants reporting that they had not previously given much thought over how their information was, and is, used by the platforms that host OHCs, it can be inferred that when participants in social media studies are asked about their privacy, they are likely to report on their social privacy concerns. Secondly, throughout the interviews when participants were asked about what they believed happens with their information, there was a shared desire to learn more about the topic. Whether participants have since taken the time to learn about online practices, through revisiting privacy policies or reading articles, is unknown.

**Limitations**

The qualitative nature of this study provides a detailed view of the way in which patients with IBD perceive social and information privacy concerns and, consequently, how they deal with them. In this paper we have inferred a low level of understanding from the use of broad terms, such as 'social media platforms collect everything'. A lack of prior thought about what platforms collect was also a common factor. Future studies can better quantify patients' understanding of the digital economy and its impact on risk perception through other methods, such as applying methods similar to the Understanding Survey by Dot Everyone (2018).

Despite members of the IBD community spanning the world, the majority of participants in this study were from the United Kingdom (n=32). Therefore, the information concerns reported may be different for a study focused specifically on the USA or South Africa, where participants in this study talked more freely about the financial cost of healthcare.

Participants for this study were self-selecting and the majority (n=36) describe themselves as non-posters, or 'lurkers' (Jenkins, Ford & Jean, 2013). Previous research indicates that individuals who self-disclose on social media represent around 1% of an online community. Furthermore, it should be noted that not all patients will engage with internet technologies for peer support, either from a lack of awareness, interest or internet connection. While previous research indicates motivations predict posting behaviours (Choi et al., 2017), future research could compare the posting behaviours between health communities and special interest groups; exploring whether the 1% rule still applies.

**Conclusion**

In this study, we explored information and social privacy concerns expressed by patients with Inflammatory Bowel Disease through a qualitative lens to better understand what risks are perceived. Our findings suggest that participants have a good understanding of social privacy concerns, identifying risks of social exclusion and career hindrances. With social privacy risk awareness, participants described strategies of audience management (through privacy settings, multiple accounts and Facebook Groups) and self-censorship. Conversely, participants struggled to identify information privacy risks beyond unsolicited advertising and demonstrated mixed understandings of data collection and processing practices by the platforms that store self-disclosed health information. The misconception of information privacy safeguarding from social privacy mitigations indicate a need for social media companies to clearly communicate how data disclosed on the platforms is processed and used. With so few individuals reading terms and conditions in this study, and previous studies (Rao et al., 2016; Debatin & Lovejoy, 2009), it is clear that the methods used do not transfer knowledge effectively enough; raising questions around true informed consent.

The participants in this study demonstrated (and were aware of) a low level of understanding of what data is collected by social media. While there is a low understanding inferred in this study, we suggest that future research is required to further understand the correlation between knowledge, information privacy concern and mitigating actions within the OHC context. Participants demonstrated an interest in increasing knowledge of their digital footprints and how they might mitigate against any risks they perceive. In order to increase knowledge and agency over personal data, future research projects should explore the creation of learning materials tailored towards health communities. Finally, though both facets of online privacy warrant continued research to improve safeguards for users, this paper indicates a need for future social media privacy research to clearly indicate which facets of online privacy they are exploring.